\documentclass[
 aps,
 prx,
 amsmath,amssymb,amsfonts,
 reprint,
 10pt,
 superscriptaddress,
 floatfix,
longbibliography,
 noeprint
]{revtex4-2}
\usepackage{xcolor}
\usepackage{amssymb}
\usepackage{mathtools}
\usepackage{natbib}
\usepackage{amsmath}
\usepackage{amsfonts}
\usepackage{comment}
\usepackage{graphicx}
\usepackage{mathrsfs}
\usepackage{dcolumn}
\usepackage{bm}
\usepackage{cancel}
\usepackage{mathbbol}
\usepackage{epsfig}
\usepackage{units}
\usepackage{esint}
\usepackage{multirow}
\usepackage{algpseudocode}
\usepackage{algorithm}
\usepackage{mathtools}
\usepackage{slashed}
\usepackage{soul}
\usepackage{mwe}
\usepackage{tikz}
\usepackage{color}
\usetikzlibrary{quantikz}
\usepackage[normalem]{ulem}

\usepackage[colorlinks,
            linkcolor=blue,    
            anchorcolor=black,
            citecolor=blue,  
            urlcolor=blue,    
 	        bookmarks=true,        
	        bookmarksopen=true,    
	        bookmarksnumbered=true,
]{hyperref}

\def\oper{{\mathchoice{\rm 1\mskip-4mu l}{\rm 1\mskip-4mu l}
{\rm 1\mskip-4.5mu l}{\rm 1\mskip-5mu l}}}

\definecolor{rot}{rgb}{0.75,0.05,0.25}
\definecolor{hellgrau}{gray}{0.5}
\definecolor{blau}{rgb}{0,0,0.7}

\makeatletter
\newcommand{\shorteq}{%
  \settowidth{\@tempdima}{-}
  \resizebox{\@tempdima}{\height}{=}%
}

\begin{document}


\title{Spectra of noisy 
parameterized quantum circuits:  Single-Ring universality}

\author{Kristian Wold}
\affiliation{Department of Computer Science, OsloMet – Oslo Metropolitan University, N-0130 Oslo, Norway}
\author{Pedro Ribeiro}
\affiliation{CeFEMA, Instituto Superior Técnico, Universidade de Lisboa, 1049-001 Lisboa, Portugal}
\author{Sergey Denisov}\email{sergiyde@oslomet.no}
\affiliation{Department of Computer Science, OsloMet – Oslo Metropolitan University, N-0130 Oslo, Norway}

\date{\today }

\begin{abstract}
Random unitaries are an important resource for quantum information processing. While their universal properties have been thoroughly analyzed, it is not known what happens to these properties when the unitaries are sampled on the present-day noisy intermediate-scale quantum 
computers. We implement parameterized circuits, which have been proposed as a means to generate random unitaries,  on 
an IBM Quantum processor
and model these implementations as quantum maps. 
To retrieve the maps, a gradient-based process tomography protocol is developed and used.  We find the spectrum of a map to be either an annulus or a disk depending on the circuit depth and detect an annulus-disk transition. 
By their spectral properties, the retrieved maps appear to be very similar to a recently introduced ensemble of random maps,  for which spectral densities can be analytically evaluated. Our results 
establish, via Dissipative Quantum Chaos theory, a connection between intrinsic properties of present-day noisy intermediate-scale quantum (NISQ) computing platforms and non-Hermitian random matrix theory.
\end{abstract}
\maketitle

\section{Introduction}
Effective methods to sample random unitaries~\cite{karol} are in high demand 
for quantum cryptography~\cite{crypto1,crypto2}, quantum information processing~\cite{info}, and algorithm design~\cite{algorithm}. From a theoretical perspective, it is natural to  sample unitaries uniformly over the unitary group~\cite{Haar}. 
However, sampling these \emph{Haar-random} unitaries on a quantum computer is impractical, requiring a number of gates that scale exponentially with the number of qubits~\cite{knill}. To address this challenge, the concept of $t$-design~\cite{tdesign} was adapted~\cite{tdesignQ} to devise shallow sampling circuits, thereby reducing the scaling to polynomial~\cite{tdesignPol,tdesignLin}.

Pseudo-random unitaries can be generated using circuits that consist of alternating layers of random single-  and two-qubit operations~\cite{emerson2003}. Recently, Sim \emph{et al.}~\cite{parameter} considered a selection of  circuit blocks consisting of layers of parameterized single- and two-quibit rotations. Unitaries are generated by concatenating $\ell$ blocks and  sampling over the rotation angles of the resulting circuit. 
By increasing $\ell$, one can quickly achieve an 'expressibility' (a practical metric for evaluating sampling quality) comparable to 
finite-size Haar-random sampling~\cite{parameter}.

Useful pseudo-random $d$-dimensional unitaries, $U^{\mathrm{PR}}_d$, 
are expected to posses key properties of their Haar-distributed counterparts. 
From the spectral perspective, this implies that eigenvalues and eigenvectors of $U^{\mathrm{PR}}_d$ should exhibit the features of the Circular Unitary Ensemble~\cite{karol, emerson2003}. In particular, in the limit $d \rightarrow \infty$, the eigenvalues  are expected to uniformly cover the unit circle while their two-point correlation function  is expected to follow the Wigner distribution for $\beta = 2$~\cite{karol,emerson2003,karol2}.

What happens to these spectral features when  $U^{\mathrm{PR}}_d$ are sampled on a present-day quantum computer? 
As implementations on a 
NISQ~\cite{Preskill} platform will inevitably be non-unitary, analyzing features of the unitary spectra is no longer possible. 
The ability to address the above question thus hinges on answering the following one: Being in the NISQ realm, where should we look for \emph{potentially} universal spectral features?

In this work,  we analyze spectra of three- and four-qubit parameterized circuits, implemented on IBM quantum processors~\cite{IBM}, by modeling these implementations
as completely-positive trace-preserving  (CPTP) maps~\cite{Choi1975}. We find that, depending on the depth of a circuit, the distribution of the eigenvalues of the corresponding map follows the 'Single Ring' scenario~\cite{ring1,ring2}, i.e., it comes either in the form of an annulus or a disc. 
We show that a recently introduced ensemble of  random maps~\cite{dunit} provides a framework for a 
classification of NISQ parameterized circuits using two quantifiers only, the strength of dissipation and its rank. Figure 1 outlines our retrieval procedure, highlighting its three main components: parameterization, tomography, and optimization. We begin by explaining the procedure and then present the experimental results.

\section{Map parametrization}
NISQ implementations of quantum circuits  can be modeled as completely-positive trace-preserving quantum (CPTP) maps [also referred to as "channels"~\cite{Wilde2017}, "operations"~\cite{Nielsen}, and "processes"~\cite{Gross2019}]. 
Some recent works have already attempted to approximate NISQ circuits and individual gates~\cite{David2021,LT2022,Bartolomeo2023} using Liouvillians~\cite{Alicki2007}, which yield specific types of CPTP maps~\cite{Wolf2008,Wolf2008_2}. 
Here we consider the most general type of CPTP maps (which we will henceforth refer to as "maps"), operating in $d=2^n$-dimensional Hilbert  space, with $n$ being the number of qubits.

\begin{figure*}[t]
\begin{center}
\includegraphics[width=\textwidth]{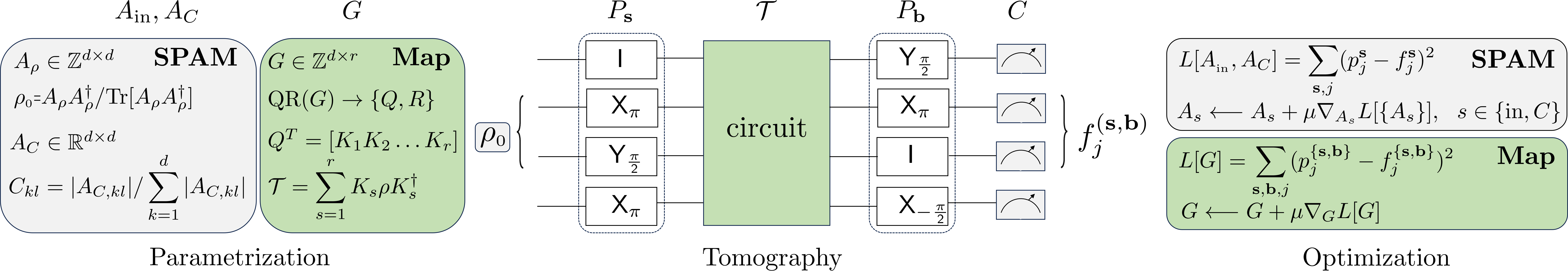}
\caption{We retrieve  map $\mathcal{T}$  
approximating a NISQ circuit with the frequencies $f^{(\bf{s},\bf{b})}_j$ collected performing tomography with Pauli strings $P_{\bf{s}}$ and $P_{\bf{b}}$. The map is parameterized with complex matrix $G$, while SPAM errors, modeled using corruption matrix $C$ and input density operator $\rho_0$, are encoded with matrices $A_{C}$ and $A_{\rho}$.
We determine parameter matrices by minimizing the quadratic loss function $L$, which is based on the differences between frequencies and probabilities yielded by the map, Eq.~(\ref{eq:prob}). To find matrices $A_{C}$ and $A_{\text{in}}$, we optimize without the circuit, setting $\mathcal{T}=id$ and $P_{\bf{b}}=\{z,z,...,z\}$.
}
\end{center}
\label{fig:1}
\end{figure*}

While a one-to-one parameterization of maps
has been proposed~\cite{parametrization1}, we find it impractical due to the trace-preservation condition, which imposes nontrivial constraints on parameter values. Instead, we introduce a parametrization based on the Kraus form~\cite{Wilde2017}, 
\begin{equation}\label{eq:map}
 \mathcal{T}_r(\rho;\boldsymbol{\theta}) = \sum_{s=1}^{r}K_s(\boldsymbol{\theta})\rho K_s(\boldsymbol{\theta})^{\dagger},~ 
    \sum_{s=1}^{r} K_s(\boldsymbol{\theta})^{\dagger}K_s(\boldsymbol{\theta}) = \oper_{d},
\end{equation}
where $\boldsymbol{\theta}$ is the parameter vector,  
$K_i(\boldsymbol{\theta}) \in \mathbb{C}^{d\times d}$ are Kraus operators and $1 \leqslant r  \leqslant d^2$ is the rank of the map. $r=1$ corresponds to a unitary map, while $r=d^2$ represents the maximal number of Kraus operators needed to specify \emph{any} map for a chosen $d$.

The second identity in Eq.~\eqref{eq:map} imposes the trace-preservation condition and implies that, by stacking matrix representations of Kraus operators, we get a rectangular $rd \times d$ matrix $Q$ with the isometry condition  $QQ^{\dagger}=\oper_{d}$.
We parameterize $Q$ using the QR-decomposition of a complex $rd \times d$ matrix $G$, $G = QR$. The decomposition is made unique by demanding that the upper triangular $d \times d$ matrix $R$ has positive diagonal elements~\cite{Haar}. This procedure yields a set of Kraus operators  parameterized by $rd^2$ complex numbers given by $\{G_{l,n}\}$.
The $2d^2$ dimensional vector $\boldsymbol{\theta}$ of  parameters is constructed by stacking the real and imaginary parts of $\{G_{l,n}\}$.

Our parameterization is only surjective due to the unitary freedom~\cite{Wilde2017}.  
From perspective of optimization, the fact that different vectors $\boldsymbol{\theta}$ yield the same set of Kraus operators -- and therefore the same map --
is not disadvantageous as both the search space and the optimal manifold dimensions are increased.

Since our retrieval procedure is based on tomography data, we need to account for state preparation and measurement (SPAM) errors. 
We do this by considering a non-ideal initial density operator $\rho_0 = A_{\rho}A_{\rho}^{\dagger}/\text{Tr}[A_{\rho}A_{\rho}^{\dagger}]$, parameterized by a complex $d \times d$ matrix $A_{\rho}$.  
As aforementioned, here the surjective nature of the parameterization is potentially advantageous for the optimization. To account for measurements errors,  generic positive operator-valued measure (POVM) sets $\{E_j\}_{j=0,\cdots, d-1}$~\cite{Wilde2017}, are conventionally used; see, e.g., Refs.~\cite{Presti2004,Lundeen2008,LT2022,Hetzel2023}.  However, in the present-day qubit transmon platforms, classical readout errors are dominant~\cite{Oszmaniec2020}, which correspond to diagonally dominant matrices $\{E_j\}$, with the magnitude of off-diagonal elements often falling below the resolution threshold. 
We find that for $n \geqslant 3$ and typical number of shots, $N_s\leqslant 2^{11}$, a full-fledged POVM modeling would be highly inefficient and even detrimental~(see Appendix B) to the accuracy of the description of measurement errors as compared to the 'diagonal' readout description~\cite{Oszmaniec2020}. Therefore, we replace the POVM set with $d \times d$ stochastic \emph{corruption matrix}
$C$, so that $(E_j)_{kl}=\delta_{kl} C_{jl}$, which is parameterized by a $d \times d$ real matrix $A_C$, $C_{jl} = |(A_\text{C})_{jl}|/\sum_{k=1}^d |(A_\text{C})_{kl}|$. In total, SPAM errors are parameterized by real  $3d^2$ vector $\boldsymbol\omega = \{ \text{Re} (A_{\rho})_{i,j}, \text{Im} (A_{\rho})_{k,l}, (A_\text{C})_{k,l} \}_{k,l=1,..,d}$.  

\section{Tomography}

Several schemes have been proposed~\cite{Nielsen1997, Lidar2008, Knee2018,Gross2019} and experimentally realized (for $n=1,2$)~\cite{Altepeter2003, Shabani2011, Rodionov2014} to retrieve maps from measurements. Many of these schemes are based on 
the state-map duality (Choi-Jamiołkowski isomorphism)~\cite{Wilde2017}, thus reducing the problem to a state tomography; some~\cite{Shabani2011,Rodionov2014,Gross2019} use the idea of compressed sensing~\cite{compressed} which allows to reduce the number of measurements in the case of low-rank maps.

Considering an arbitrary input state and/or measurement basis 
is infeasible as the preparation of such state would require a dedicated 
parameterized circuit of growing depth (in the worst case, exponentially in $n$), thus leading to an exacerbation of the original problem. We therefore utilize the idea of the Pauli string - based tomography~\cite{Gross2019,LT2022}, that uses only product input states, $\vert {\boldsymbol s} \rangle = \bigotimes_{i=1}^n \vert s_i \rangle$, with each qubit being in one of the six basis eigenstates,
$\vert s_i \rangle \in \lbrace \vert \pm x_i \rangle, \vert \pm y_i \rangle, \vert \pm z_i \rangle \rbrace$. The input product state is prepared with a layer of single-qubit rotations; see Fig.~1, panel "Tomography". After passing the circuit, qubits are measured in different Pauli bases, $b_j \in  \lbrace x_j,y_j,z_j \rbrace$.  A single Pauli measurement mode is specified by vector ${\boldsymbol p}=(\boldsymbol{s},\boldsymbol{b})$,  ${\boldsymbol s} = \lbrace s_1,..s_n \rbrace$ and ${\boldsymbol b}= \lbrace b_1,..b_n \rbrace$. The total number of measurement modes is thus $18^n$. 

While tomography limited to Pauli strings has not been proven to be able to retrieve {\it any} map to arbitrary precision~\cite{Gross2019}, in practice, its accuracy is further limited by the number $N_{\text{mode}}$ of measurement modes used, $N_m < 18^n$, and by the number of shots per mode, $N_s$, performed to estimate the  probability distributions on the binary string outputs. However,  as we discuss below, even under these restrictions, the Pauli-string tomography demonstrates the capability to retrieve spectral densities of the test maps with high accuracy.

\section{Optimization}

Similar to Ref.~\cite{Kockum2023}, 
we use a quadratic cost function 
$
L(\boldsymbol{\theta}) = \sum_{\bf{s},\bf{b},j} 
\left[ \hat{p}^{(\bf{s},\bf{b})}_j(\boldsymbol{\theta}) - f^{(\bf{s},\bf{b})}_j \right]^2$,
where
$f^{(\bf{s},\bf{b})}_j$ is the frequency of 
the $j$-th binary output obtained in the experiment with Pauli modes $P_{\bf{s}}$ and $P_{\bf{b}}$, and  \begin{equation}\label{eq:prob}
\hat{p}_j^{\bf{s,b}}(\boldsymbol{\theta})= \sum_{l=1}^{d} C_{jl} \cdot ( P_{\bf{b}}\mathcal{T}(P_{\bf{s}}\rho_{\text{\tiny{0}}} P^{\dag}_{\bf {s}};\boldsymbol{\theta})  P^{\dag}_{\bf{b}})_{ll}
\end{equation}
is the probability yielded by the map, Eq.~(\ref{eq:map}), filtered through the corruption matrix. Here $(A)_{lk}$ is the   element of operator $A$ expressed in the computation basis.

To search for minima of the cost function,
we employ the Adam optimizer version~\cite{kingma2017adam} of the steepest-descent  method,
$\boldsymbol{\theta} \leftarrow  \boldsymbol{\theta} - \mu \nabla_{\boldsymbol{\theta}} L(\boldsymbol{\theta})$. Note that, unlike the optimization approaches in Refs.~\cite{Guta2022,Kockum2023}, we do not need to constrain the search or project its results on a physically meaningful subspace as our parameterization guarantees that any vector $\boldsymbol{\theta}$  yields a valid trace-preserving map.

\begin{figure}[t]
\begin{center}
\includegraphics[width=1.00\columnwidth]
{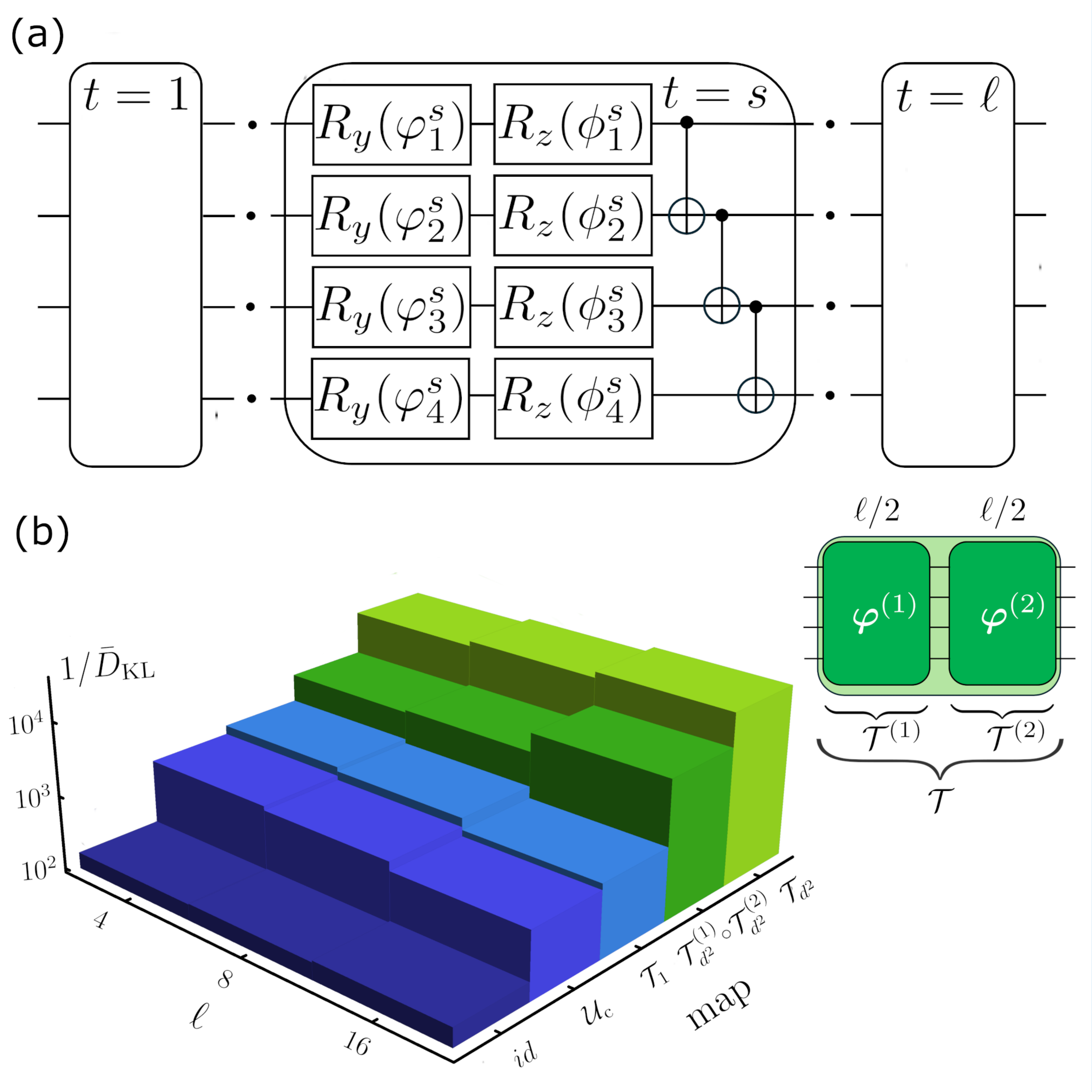}
\end{center}
\caption{(a) In the experiment we implement $n$-qubit circuits parameterized with $2n \cdot \ell$ angles, where $\ell$ is the circuit depth. Here a circuit for $n=4$ is shown.
(b) Inverse average Kullback-Leibler  divergence between frequencies obtained with three-qubit circuits, implemented on ibmq\_belem, and probabilities yielded by the identity map $id$ and retrieved rank-$r$ maps, $\mathcal{T}_r$, plotted against circuit depth $\ell$. 
Map $\mathcal{T}^{(1)}_{d^2} \circ \mathcal{T}^{(2)}_{d^2}$ results from concatenating two maps,  independently retrieved with two halves of the original circuit.}
\label{fig:2}
\end{figure}

The optimization is first performed for the SPAM parameter vector $\boldsymbol{\omega}$, without the circuit,  $\mathcal{T} = id$,  and by omitting  base rotation before the measurement, $\vec{b} = \{z,z,...,z\}$. After retrieving $\rho_0$ and the corruption  matrix $C$, we run the optimization to retrieve the map, by finding 
parameter vector $\boldsymbol{\theta}$ that minimizes $L(\boldsymbol{\theta})$.

We benchmark the  procedure 
using different maps, including exponents of Liovillians, and an experiment-relevant SPAM error model~(see Appendix C). Synthetic data sets are generated by performing $N_s= 2^{10}$ 'shots' per measurement mode (as in the experiment). Across all tested models, our procedure consistently retrieves maps with spectral supports closely resembling those of the originals.

\begin{figure*} [t]
\begin{center}
\includegraphics[width=1.98\columnwidth]{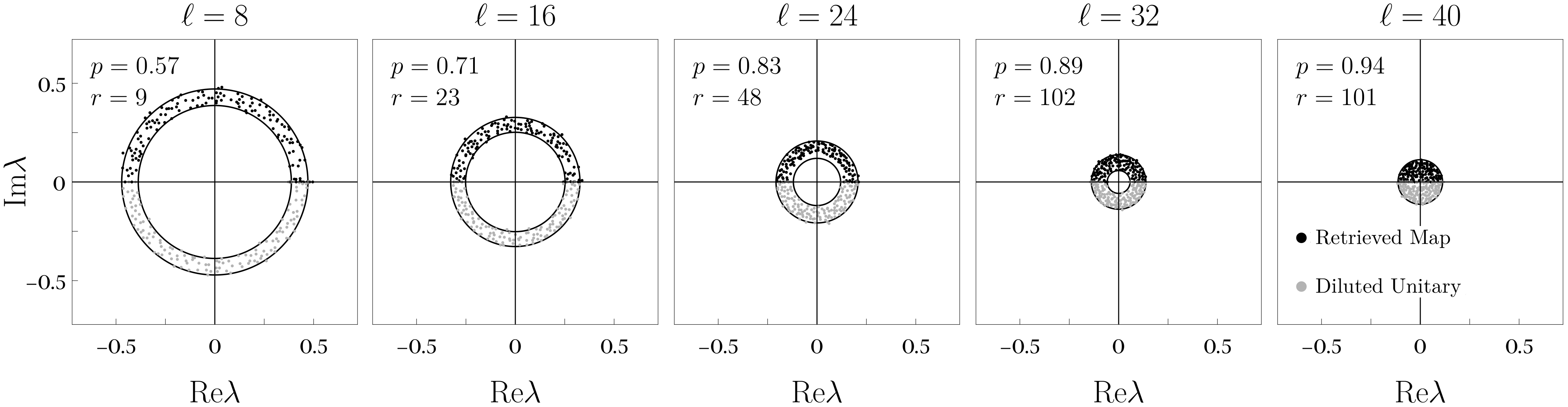}
\end{center}
\caption{Eigenvalues $\lbrace \lambda_j \rbrace,~j=2,...,d^2$ of the 
maps $\mathcal{T}_{d^2}$ retrieved with four-qubit parameterized circuits of varying depth $\ell$ implemented on ibmq\_belem. The spectra are invariant under the complex conjugation and only the eigenvalues  with $\text{Im} \lambda \geqslant 0$ 
are plotted. The eigenvalues with $\text{Im} \lambda^{\text{DQ}} \leqslant 0$, 
of the diluted unitaries $\mathcal{DU}_d(p,r)$, obtained as spectral approximations of the retieved maps, are shown together with the corresponding $p$ and $r$ values. Radii of  the outer, $R_+$, and inner, $R_-$, circles are given by $R_\pm =\sqrt{(1-p)^2 \pm p^2/r}$. For $\ell = 40$, the eigenvalues of the corresponding diluted unitaries filled a disc. 
}
\label{fig:3}
\end{figure*}

\section{Results}

For the experiment, we adapt one of the parameterized circuits considered in Ref.~\cite{parameter}; see Fig.~2(a). It is composed of $\ell$ blocks, each one consisting  of two layers of single-qubit rotations, parameterized with $2n$ angles $\lbrace \varphi_l, \phi_l \rbrace$, $l=1,...,n$, and a ladder of cNOT gates. 
This particular design allows to quickly reach high expressibility values~\cite{parameter} by gradually increasing $\ell$. For example, when limited to a number of samples $\leqslant 10^5$, the expressibility of a four-qubit parameterized circuit of depth $\ell \geqslant 4$ is indistinguishable from that obtained with the Haar-random sampling~(see Appendix A).

We run experiments with parameterized circuits implemented on the IBM transmon platform ibmq\_belem~\cite{IBM}, with $N_{\text{mode}}= 1784$ ($n=3$) and $N_m= 8704$ ($n=4$) measurement modes respectively and $N_s= 2^{10}$ shots per mode.  Figure 2(b) illustrates the significance of modeling the implemented circuits using maps rather than unitaries. The accuracy of the modeling of 
three-qubit circuits is quantified as  the  Kullback-Leibler (KL) divergence $\overline{D}_{KL}$ between the  frequencies estimated in the experiment and the probabilities yielded by the retrieved map. 
From  randomly selected $N_m= 1784$ measurement modes,
we allocate $90\%$ for training and the $10\%$ for estimating the average $D_\text{KL}$. The KL-divergence is additionally averaged over ten realizations (samples of angles). 

Although the optimized unitary map $\mathcal{T}_{r=1}$ scores slightly better than the ideal unitary map of the circuit, $\mathcal{U}_c(\cdot) = U_{c} \cdot U_{c}^{\dagger}$, the full-rank  map $\mathcal{T}_{d^2}$ yields an accuracy improvement of more than an order of magnitude. Independently optimizing maps $T^{(1)/(2)}_{d^2}$ for sub-circuits of length $\ell/2$ and concatenating the results, we obtain a map that, while yielding lower accuracy than $\mathcal{T}{d^2}$, still demonstrates substantially higher accuracy than the unitary approximation.

Figure~3 presents the spectra $\lbrace \lambda_j \rbrace,~j=2,...,d^2$ (the leading eigenvalue $\lambda_1 = 1$ is not shown) of the retrieved full-rank maps 
for random four-qubit parameterized circuits of increasing depths (the results for three-quibit circuits are presented in Appendix D). 
The eigenvalues are located inside the unit disk thus highlighting a strong non-unitarity character of the maps. Eigenvalues are arranged in an annulus-like spectral distribution whose outer and inner radii diminishing as $\ell$ increases, eventually undergoing an annulus-disc transition.

\begin{figure}[b]
\begin{center}
\includegraphics[width=0.99\columnwidth]{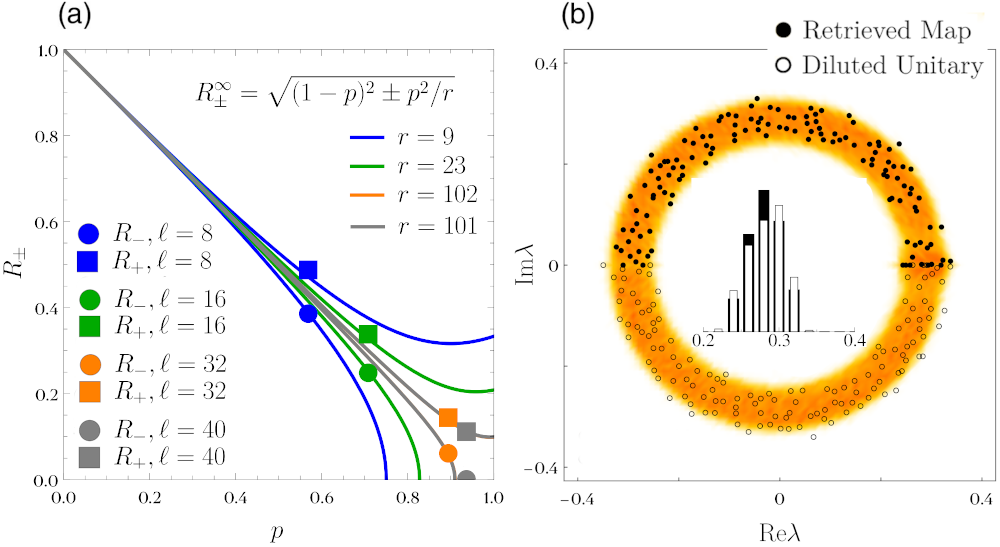}
\end{center}
\caption{(a) Outer and inner radii of the diluted unitaries  in the limit $d \rightarrow \infty$ (lines)
compared to the maximal and minimal eigenvalues $\vert \lambda_j \vert_{j=2,...,d^2}$ (symbols) of the retrieved four-qubit maps. 
(b) Spectra of the map retrieved for a four-qubit map of the depth $\ell = 16$ and of a randomly sampled diluted unitary from the ensemble $\mathcal{DU}_{0.71,23}$. The eigenvalues  are superimposed onto the spectral density obtained by sampling over $10^4$ diluted unitaries  from $\mathcal{DU}_{0.71,23}$. The inset compares histograms of $\vert \lambda_j \vert$ and $\vert \lambda^{\text{DU}}_j \vert$.}
\label{fig:4}
\end{figure}

A similar scenario was recently reported for the ensemble of random maps~\cite{dunit},
\begin{equation}\label{eq:DU}
\mathcal{DU}_{p,r}(\cdot) = p U\cdot U^\dagger + (1-p) \sum_{s=1}^r K_s \cdot K_s^\dagger, 
\end{equation}
which we henceforth refer to as the \emph{diluted unitaries} (DU). A DU ensemble is parameterized by the relative strength of the dissipative term, $0< p <1 $, and its rank, $r= 1,..., d^2$. $U$'s are sampled from the ensemble of  Haar random unitaries while the Kraus operators $\{K_s\}$ are sampled by using rectangular $d \times d r$ Ginibre matrix (as described in the  "Parameterization" section). 
Because of the close similarity between the spectra retrieved in the experiment and those obtained for DU ensembles, we employ the latter as a tool to describe and analyze how the spectral densities of the NISQ circuits evolve with increasing $\ell$.

To identify ensemble $\mathcal{DU}_{p,r}$ with the spectral support best approximating that of the retrieved map, we minimize the $2$-Wasserstein distance  between the Gaussian kernel density estimators of the eigenvalues of the retrieved map and those of a typical member of $\mathcal{DU}_{p,r}$~(see Appendix C).

It is noteworthy that, although we used a generic full-rank map, Eq.~(1), to model the action of the NISQ circuits, the corresponding approximating DUs have ranks substantially lower than $d^2$ which, in addition, depend on $\ell$. Thus, together with the dissipation strength $p$, the rank of the approximating DU can be used to label an ensemble of parameterized NISQ circuits of fixed depth $\ell$. The values of $p$ and $r$ for different $\ell$ are presented in Fig.~3 as well as  the spectrum of a randomly sampled member of $\mathcal{DU}_{p,r}$.

In the asymptotic limit, $d \to \infty$, the spectral densities of DUs can be evaluated by using free probability theory~\cite{FP}. In this limit, the inner $R_-$ and outer $R_+$ radii of the spectral support are given by $R_\pm =\sqrt{(1-p)^2 \pm p^2/r}$~\cite{dunit}.
Figure~\ref{fig:4}(a) presents a comparison of the asymptotic radii  as  functions of $p$ for different $r$, with the smallest and largest absolute values in the spectrum  $\lbrace \lambda_j \rbrace_{j=2,...,d^2}$ of the retrieved map. The matching is quite accurate, even for a relatively small  dimension $d=2^4=16$.  Figure~4(b) shows that not only the spectral support matches, but even the distribution of the eigenvalues  is well reproduced by that of the approximating DU. Note that the eigenvalues displayed in the lower semi-plane of Fig.~4(b) correspond to a random sample from the ensemble $\mathcal{DU}_{0.71,23}$.

\section{Conclusion}
We demonstrated that NISQ parameterized circuits, when  modeled as quantum maps, display universal spectral properties.
Specifically, their spectral supports, which come in the form of an annulus or a disc, are independent of the particular parameter choices and only depend  on the circuit depth. 

We found that the Diluted Unitary model~\cite{dunit} is able to accurately reproduce the evolution of the spectral support with the circuit depth~\cite{dunit}.
This  suggests that the maps retrieved from NISQ parameterized circuits and diluted unitaries are particular manifestations of a broader underlying universality.  In Ref.~\cite{Campo2023}, the annulus-disc transition is detected in the spectra of concatenations of unitary and purely dissipative maps. Moreover, this transition is also observed in the spectra of maps obtained by exponentiating Lindbadians with random Hamiltonan and dissipative components~\cite{WRD2024}. We believe that the annulus-disc scenario can be modeled in the framework of non-Hermitian random matrix theory~\cite{NonHerm1} and can thus be related to the Single-Ring Theorem~\cite{ring1,ring2}. 


We also found  evidence that the evolution of multi-qubit states in NISQ circuits is significantly influenced  by non-Markovian effects~\cite{NonMarkovian1,NonMarkovian2}. Specifically, while the concatenation of two maps, each one modeling a half of a NISQ circuit, offers a notably improved approximation of the entire circuit compared to the unitary description (see Fig.~2b), the concatenated map exhibits significantly reduced accuracy compared to the map approximating the whole circuit. Quantifying the degree of non-Markovianity~\cite{NonMarkovian1,NonMarkovian2} in NISQ  circuits is an interesting task, notwithstanding experimental challenges.

Finally, our findings emphasize present-day NISQ computers as versatile platforms for exploring Dissipative Quantum Chaos (DQS). This emerging theory aims, by utilizing the theoretical framework of non-Hermitian random matrix theory~\cite{NonHerm1}, to figure out how properties of open quantum evolution, induced by CPTP maps and quantum Liouvillians, are encoded in the spectral features of the latter~\cite{Karol2008, Denisov2019, Ca19, COOG19, SRP20, Luitz2020, dunit, David2021, Tarnowsli2021, Campo2023,Sa2023,Kawabata2023,Orgad2024,Campo2024}. In this perspective, the existing NISQ platforms can play a role similar to the role microwave billiards  played in the experimental verification of the Hamiltonian Quantum Chaos theory~\cite{Stock1999}.

\section{Acknowledgements}
We thank Karol \.Zyczkowski, Toma\v{z} Prosen, Dariusz Chru{\'s}ci{\'n}ski, and Lucas S{\'a} for useful discussions. We acknowledge the use of IBM Quantum services for this work. This research is supported by Research Council of Norway, project “IKTPLUSS-IKT og digital innovasjon
- 333979” (KW and SD) and by FCT-Portugal 
(PR), as part of the QuantERA II project “DQUANT: A Dissipative Quantum Chaos perspective on Near-Term Quantum Computing” via Grant Agreement No. 101017733. PR acknowledges further support from FCT-Portugal through Grant No. UID/CTM/04540/2020. 

\appendix
\section{The parameterized circuit and its expressibility}
\label{app:parametrized}

For the experiment, we adapted a circuit from Ref.~\cite{parameter}, referred therein as 'Circuit 2'. A  circuit block, sketched in Fig.~2a in the main text, consists of two layers of single-qubit parametrized rotations and a ladder-like chain of CNOT gates (we replaced \(R_x\) rotations in the first layer of the original design with \(R_y\)). The block is parameterized with $2n$ angles, $\lbrace  \varphi_1,...,\varphi_n, \phi_1,...,\phi_n \rbrace$, where $n$ is the number of qubits.

The expressibility~\cite{parameter} of the circuit $U_{\ell}(\boldsymbol{\psi})$, consisting of  $\ell$ blocks and parameterized by $2n\ell$-dimensional vector of angles,  $\boldsymbol{\psi}$, is defined as the Kullback-Leibler (KL) divergence,
\begin{equation}\label{eq:map1}
\text{Expr}(N_{\text{sample}}) = D_{\text{KL}} \bigl[ p_{\ell}(F;N_{\text{sample}})\vert p_{\text{Haar}}(F;N_{\text{sample}})\bigr],
\end{equation}
between the distribution of the fidelitiy $p_{\ell}(F;N_{\text{sample}})$,  $F = \left\vert \langle \boldsymbol{0}\vert U_{\ell}(\boldsymbol{\psi})\vert \boldsymbol{0} \rangle\right\vert^2$, obtained with the circuit by uniformly sampling $N_{\text{sample}}$ times over $2n\ell$ angles, and the distribution corresponding to the Haar random sampling, $p_{\text{Haar}}(F;N_{\text{sample}})$ (note that, in the limit $N_{\text{sample}} \to \infty
$, the analytical form of the distribution is known,
$p_{\text{Haar}}=(d-1)(1-F)^{d-2}$~[S2]). Finally, as
the probability distributions need to be estimated with histograms, there is one more parameter which is the number $N_{\text{bins}}$ of bins per unit interval. We use the same number, $N_{\text{bins}}=75$, as in Ref.~\cite{parameter}. 

\begin{figure}[t]
\begin{center}
\includegraphics[width=0.95\columnwidth]{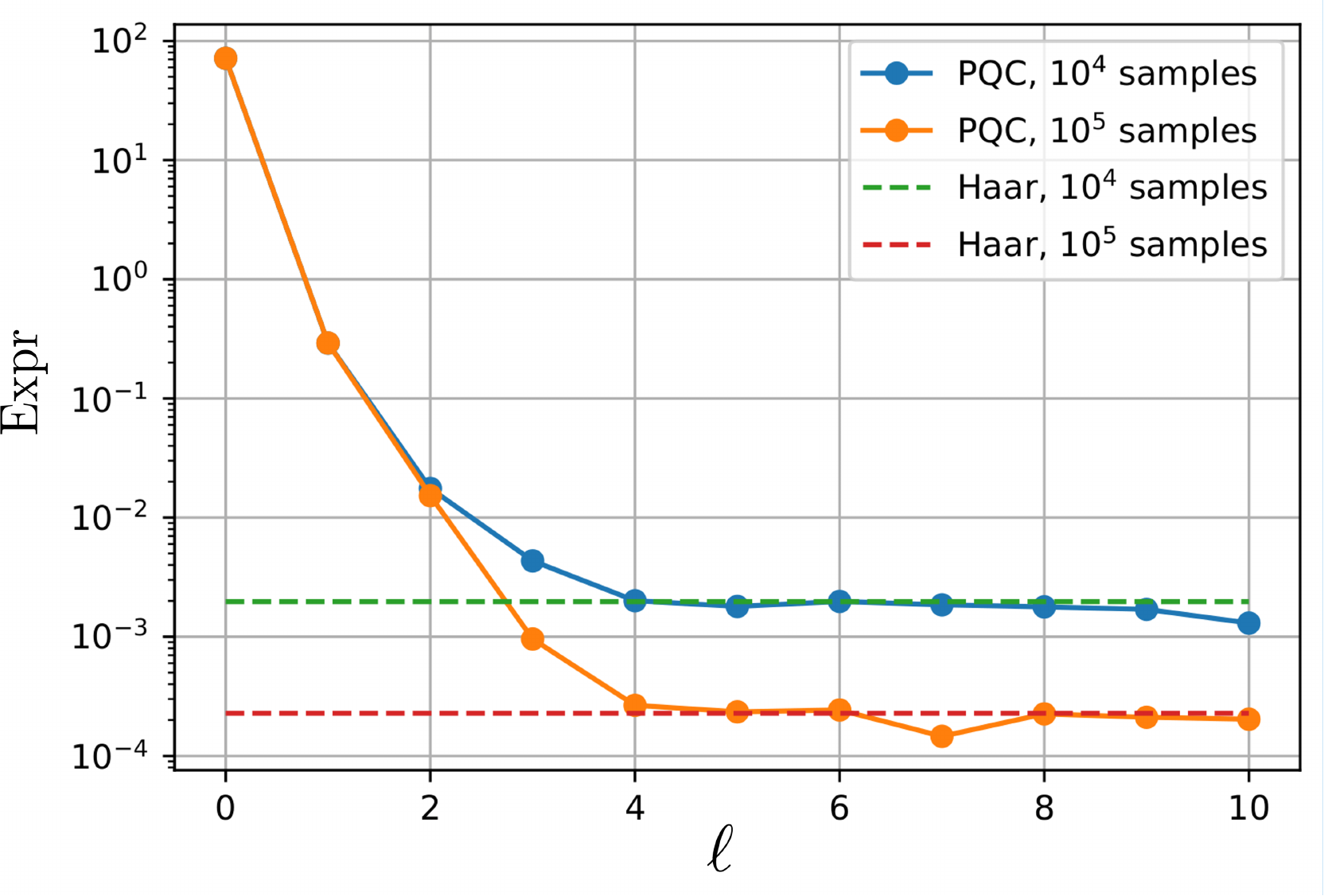}
\end{center}
\caption{Expressibility, Eq.~\ref{eq:map1},  of the parameterized circuit used in the experiment, as function of the number of concatenated blocks, $\ell$, for two numbers of samples. The number of bins per unit interval is 75.}
\label{PQC_Haar_approx_4qubits}
\end{figure}

In Fig.~\ref{PQC_Haar_approx_4qubits}, we show the expressibility of the circuit as the function of its depth $\ell$, for two numbers of $N_{\text{sample}}$. Starting from $\ell = 4$, the expressibility of the circuit matches the expressiblity of the genuine Haar-random sampling for $N_\text{sample} = 10^4$ and $10^5$.

\section{POVM set and corruption matrix}

We parameterize the POVM set, $\{E_j\}$, by using Ginibre matrices $G_j = A_j + iB_j$ ($A_j,B_j \in \mathbb{R}^{d\times d}$),
\begin{equation}\label{eq:povm}
    E_j = \frac{1}{\sqrt{D}}H_j\frac{1}{\sqrt{D}},
\end{equation}
where $H_j = G_jG_j^{\dagger}$, and $D = \sum_{i=1}^d{H_i}$ is the normalizing matrix. 

We restrict the measurement errors to classical read-out errors and introduce \emph{corruption matrix} $C$, a stochastic matrix which is parameterized in the following way:
Let $A_1 \in \mathbb{R}^{d\times d}$. Then,
\begin{equation}
    C_{ij} = \frac{|(A_1)_{ij}|}{\sum_{k=1}^d|(A_1)_{kj}|}.
\end{equation} The corruption matrix can be related to the  POVM set $\{E_i\}$ by setting all the off-diagonal of all $E_j$'s to zero and defining elements of the corruption matrix as
\begin{equation}
     C_{ik} \coloneqq  (E_i)_{kk}.
\end{equation}

The off-diagonal elements of $E_j$ have absolute values substantially lower than diagonal elements. Thus finite-size fluctuations in the estimation of the off-diagonal elements can accumulate, leading to changes comparable to (or even exceeding)  those induced by the statistical fluctuations of the diagonal elements. In this scenario, accurate estimation of the non-diagonal elements is statistically challenging and rather useless as the main contribution to errors comes from the diagonal elements.
This motivated us to use the corruption matrix instead of a generic POVM. A further advantage is that the former has $\mathcal{O}(d^2)$ free parameters as opposed to $\mathcal{O}(d^3)$ of the latter.

By using synthetic benchmarks (as described in the next session), we find that the limitations in the number of measurement modes, $N_m$, and shots, $N_s$, emerging in the experiment, do not allow us to collect sufficient data to ensure that the retrieved  POVMs surpass the corruption matrix in modeling measurement errors. Instead, the POVM approach exhibits inferior performance, making the corruption matrix the preferable option; see  Appendix B.

\section{Benchmarking}

To benchmark the procedures for the retrieval of SPAM models and maps, we generate synthetic data that mimic the experimental situation.

\subsection{SPAM Model}

We define the synthetic SPAM model as
\begin{align}\label{eq:synthetic_spam}
\begin{split}
    \rho_{in} &= c_1\ket{\boldsymbol{0}}\bra{\boldsymbol{0}}
               + (1-c_1)\,\delta\rho,\\
    E^{\text{true}}_{j}
              &= c_2\ket{\boldsymbol{j}}\bra{\boldsymbol{j}}
               + (1-c_2)\,\delta E_{j}.
\end{split}
\end{align}
Here, $c_1, c_2 \in [0,1]$ and $\ket{\boldsymbol{0}}\bra{\boldsymbol{0}}$ and $\ket{\boldsymbol{j}}\bra{\boldsymbol{j}}$ represent the density operator corresponding to the ideal initial state and projectors on the basis states. We mixed these operators, in the convex way, with randomly sampled quantum states~$\delta \rho$ and POVM elements~$\delta E_j$, respectively. Using them, we calculate the probabilities $\{ p_{j}^{\boldsymbol{s}} \}$ over binary string outputs $\{\boldsymbol{j}\}$ by applying different Pauli strings to the initial state and “measuring” the results with the POVMs,

\begin{equation}\label{eq:spam_expectation_value}
  p_{j}^{\boldsymbol{s}}
  = \mathrm{Tr}\bigl[E^{\text{true}}_{j}\,P_{\boldsymbol{s}}\,
                  \rho_{in}\,P_{\boldsymbol{s}}^{\dagger}\bigr].
\end{equation}

Finally, we obtain synthetic data—“frequencies” $\hat{f}_{j}^{\boldsymbol{s}}$—by sampling from the distribution in Eq.~\eqref{eq:spam_expectation_value} $N_{\text{sample}}$ times.

\begin{figure}[t]
\centering
\includegraphics[width=0.99\columnwidth]{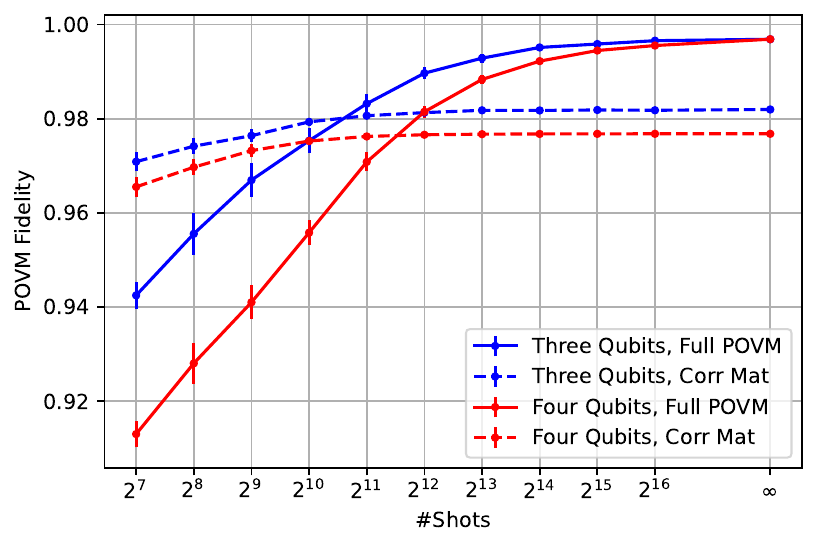}
\caption{POVM Fidelity, Eq.~\eqref{eq:PF}, between ground truth and retrieved SPAM models fitted to synthetic data, as a function of the number of shots. Both POVM sets (solid line) and corruption matrices (dashed line) were optimized.}
\label{fig:S2}
\end{figure}

\begin{figure*}[t]
\centering
\includegraphics[width=1.8\columnwidth]{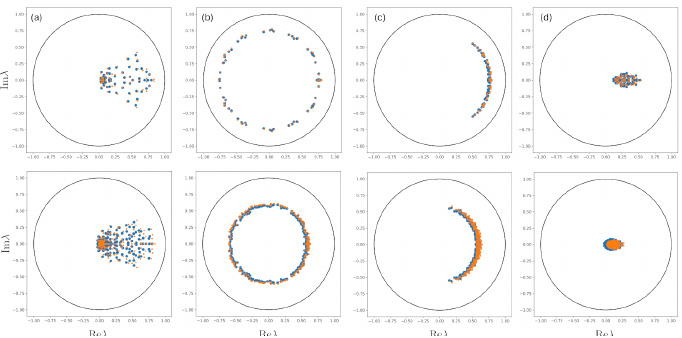}
\caption{Spectra of original maps (blue dots), Eqs.~\eqref{eq:synthetic_spam},\eqref{eq:spam_expectation_value}, and of the retrieved maps (orange dots) for $n=3$ (upper panel) and $n=4$ (lower panel). Parameters of the map are (a) $\alpha = 1$, $\beta = 0.1$, $r = 1$, (b) $\alpha = 10^{4}$, $\beta = 10^{-3}$, $r = 16$, (c) $\alpha = 10^{2}$, $\beta = 10^{-3}$, $r = 16$, (d) $\alpha = 1$, $\beta = 10^{-2}$, $r = 8$.}
\label{fig:S3}
\end{figure*}

For the benchmarking, we use $c_1 = 0.9$ and $c_2 = 0.8$, to emulate $10\%$ and $20\%$ corruption of the initial state and the classical read-out errors, respectively. We consequently run two optimization procedures, for the corruption matrix and for the POVM set, with the same synthetic data, obtained for $N_m= 6^n$ randomly selected measurement modes and different $N_{\text{sample}}$. The retrieved POVM set $\{E_j\}$ is then compared to the ground-truth set $\{E^{\text{true}}_{j}\}$, by calculating the POVM fidelity, 
\begin{equation}\label{eq:PF}
    {\text{Fidelity}}(\{E^{\text{true}}_{j}\}, \{E_j\}) = \frac{1}{d}\sum_{i=1}^d {\text{Tr}}[\sqrt{E^{\text{true}}_{j} \cdot E_j}] \in [0,1].
\end{equation}
Figure~\ref{fig:S2} presents the results of the test.

When optimizing SPAM error models, we observed that multiple choices of $\rho_0$ and ${E_i}$  result in local minima of the cost function which yield the same values for the latter. This degeneracies arise from permutations of the computational basis so that each set of operators corresponds to a physically distinct SPAM error model. The problem can be resolved by identifying the relevant solution as the one that has $\rho_{in}$  closest to the ideal initial state, $\ket{\boldsymbol{0}}\bra{\boldsymbol{0}}$, the POVM elements closest to the projectors, and the corruption matrix  located near the identity $\oper_{d}$.

\subsection{Synthetic data from CPTP maps}

We benchmark our retrieval procedure using synthetic data generated with various maps and the SPAM model (describe in the previous section), under condition of fixed number of samples, $N_{\text{sample}} = 2^{10}$ (the number of shots per measuring mode used in the experiment).

Here we present the results obtained with  maps that are constructed by exponentiating different Lindbladians, $\Phi = e^{\beta \mathcal{L}}$, where $\beta \geq 0$ determines  the 'strength' of the map (e.g., $\beta=0$ yields the identity map).

The Lindbladians are constructed by combining  random Hamiltonians and random dissipative parts. Specifically, Hamiltonians are sampled as GUE matrices, $H = \frac{G_H + G_H^{\dagger}}{2}$, where $G_H \in \mathbb{C}^{d \times d}$. The dissipative part is specified with a Choi matrix  $\phi = G_{\phi}G_{\phi}^{\dagger}$, where $G_{\phi} \in \mathbb{C}^{d^2 \times r}$ and $r$ is the rank. We define the Lindbladian as a matrix,
\begin{equation}
    \mathcal{L} = -i\alpha(I \otimes H - H^{T} \otimes I) + \phi -\frac{1}{2}({\text{Tr}}_B(\phi)^{T}\otimes I - I \otimes {\text{Tr}}_B(\phi)),
\end{equation}
where $\alpha \geq 0$ determines relative contribution of the  Hamiltonian (unitary) component on the overall dynamics.

\begin{figure*}[t]
\begin{center}
\includegraphics[width=1.9\columnwidth]
{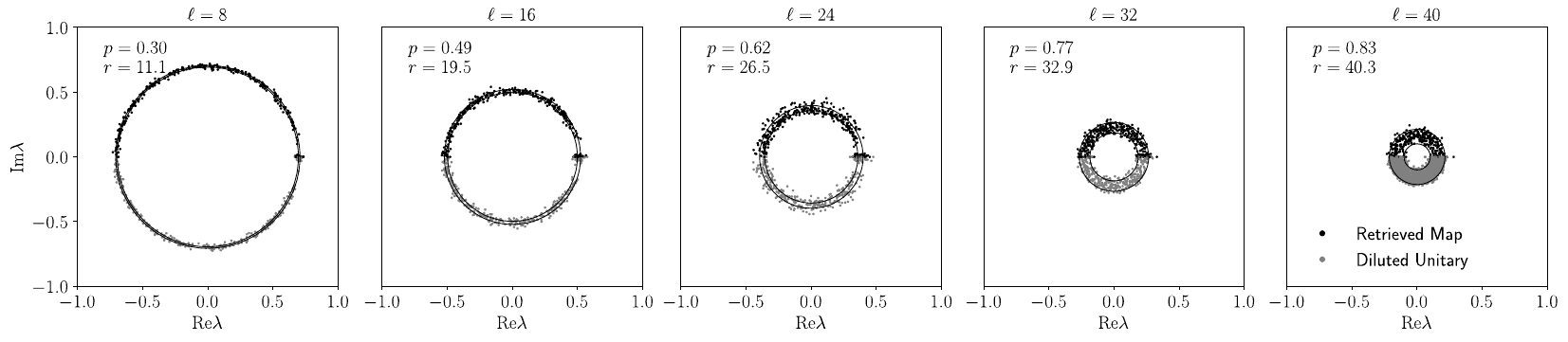}
\end{center}            
\caption{Eigenvalues $\lbrace \lambda_j \rbrace,~j=2,...,d^2$ of the 
maps $\mathcal{T}_{d^2}$ retrieved with three-qubit parameterized circuits of varying depth $\ell$ implemented on ibmq\_belem For each depth $l$, ten independent circuits was implemented and retrieved, and the resulting spectra overlaid. The spectra are invariant under the complex conjugation and only the eigenvalues  with $\text{Im} \lambda \geqslant 0$ 
are plotted. The eigenvalues with $\text{Im} \lambda^{\text{DQ}} \leqslant 0$, 
of the diluted unitaries $\mathcal{DU}_d(p,r)$, obtained as spectral approximations of the retrieved maps, are shown together with the corresponding average $p$ and $r$ values. Radii of the outer, $R_+$, and inner, $R_-$, circles are given by $R_\pm =\sqrt{(1-p)^2 \pm p^2/r}$, averaged over the ten realizations. For $\ell = 40$, the eigenvalues of the corresponding diluted unitaries filled a disc.}   \label{fig:S4}
\end{figure*}

The probability distribution 
$p_{j}^{\boldsymbol{s},\boldsymbol{b}}$ is obtained  by applying  Pauli string $P_{\boldsymbol{s}}$ to the initial state $\rho_{in}$, which is then acted upon  with the  map $\Phi$, rotated to the measurement  basis with Pauli string $P_{\boldsymbol{b}}$ and finally 'measured' by using POVM set $\{E_j\}$,
\begin{align}\label{eq:spam expectation value_2}
\begin{split}
    \rho^{\boldsymbol{s}} = 
    \Phi(P_{\boldsymbol{s}} \rho_{in} P_{\boldsymbol{s}}^{\dagger})\\
    p_{j}^{(\boldsymbol{s},\boldsymbol{b})} = 
    \mathrm{Tr}[E_{j} P_{\boldsymbol{b}} \rho^{\boldsymbol{s}} P_{\boldsymbol{b}}^{\dagger}].
\end{split}
\end{align}

Finally,  we derive synthetic data, 'frequencies'  $f_{j}^{\boldsymbol{s},\boldsymbol{b}}$, by sampling from $p_{j}^{\boldsymbol{s},\boldsymbol{b}}$ $N_{\text{shot}}$ times. We used the same number of randomly selected Pauli modes $(\boldsymbol{s},\boldsymbol{b})$ as in the experiment, $N_m= 1784$ (three qubits) and $N_m= 8704$ (four qubits).

In Figure~\ref{fig:S3}, we compare the retrieved spectra (orange dots) with the spectra of the original maps (blue dots). Although the accuracy in matching eigenvalues as a finite set of points on the plane leaves room for further improvement, the procedure is able to capture the spectral support and spectral density quite accurately.

\subsection{Spectral Fitting}

Diluted Unitaries (DU)~\cite{dunit} constitute an ensemble of random maps,
\begin{equation}
    T_{\text{Diluted}}(\rho) = (1-p)U\rho U^{\dagger} + p\sum_{j=1}^r K_j \rho K_j^{\dagger},
\end{equation}
where $U$ is a Haar-random unitary and $\lbrace K_j \rbrace_{j=1,2,...,r}$ is a set of random Kraus operators of given rank $r$. The parameter $p$ weights the relative strength of the dissipative part.

We aim to determine values of $p$ and $r$ such that the spectrum of the diluted unitary maximally overlaps with the spectrum of a given map. To measure this overlap, we introduce the spectral distance (SD) defined in the following way: 
Given two spectra, $S^{(a)} = \{\lambda^{(a)}_1, \lambda^{(a)}_2, \cdots, \lambda^{(a)}_{d^2} \}$ and $S^{(b)} = \{\lambda^{(b)}_1, \lambda^{(b)}_2, \cdots, \lambda^{(b)}_{d^2} \}$, the spectral distance (SD) is calculated as
\begin{equation}\label{eq:Spectral Distance}
    \text{SD}(S^{(a)}, S^{(b)}) = \int [p(\lambda; S^{(a)}, \sigma) - p(\lambda; S^{(b)}, \sigma)]^2 d\lambda,
\end{equation}
where $p(\lambda; S^{(a)}, \sigma)$ is a kernel density estimation over the spectrum $S^{(a)}$, using a two-dimensional Gaussian kernel of width $\sigma$. It is given by the sum
\begin{equation}
    p(\lambda; S^{(a)}, \sigma) = \frac{1}{d^2} \sum_{i=1}^{d^2} \mathcal{N}(\lambda; \lambda^{(a)}_i, \sigma) .
\end{equation}
The integral in Eq.~\ref{eq:Spectral Distance} can be computed analytically, resulting in the expression
\begin{widetext}
\begin{equation}
    \text{SD}(S^{(a)}, S^{(b)}) = \frac{1}{d^4} \sum_{i=1, j=1}^{d^2} \left[ \mathcal{N}(\lambda^{(a)}_i; \lambda^{(a)}_j, \sigma) + \mathcal{N}(\lambda^{(b)}_i; \lambda^{(b)}_j, \sigma) - 2\mathcal{N}(\lambda^{(a)}_i; \lambda^{(b)}_j, \sigma) \right].
\end{equation}
\end{widetext}

To compute the spectral distance (SD) between the spectrum $S^{(a)}$ of the map and the spectrum $S^{(b)}$ of the diluted unitary (DU), we first need to specify $\sigma$. We choose this parameter to be equal to the average nearest-neighbor distance of the eigenvalues of the map. 
By minimizing the SD, we observe that weak variations of $\sigma$ around this value result in near the same values of $p$ and $r$, thus confirming that the average nearest-neighbor distance is a reasonable choice for $\sigma$.

\section{Results for parameterized circuit with three qubits}

Figure~\ref{fig:S4}, similar to Fig.~\ref{fig:3} in the main text, presents the spectra retrieved from experimental data collected with three-qubit parameterized circuits of varying depth $\ell$, implemented on ibm\_belem. For each depth value, we sampled ten random sets of angles and retrieved the ten corresponding spectra. The spectra are plotted together with the spectra of diluted unitaries with $p$ and $r$ values obtained by minimizing the spectral distance to each of the ten circuits.

\bibliography{NISQ_spectra}

\end{document}